\documentclass[twocolumn,amsmath,amssymb,aps,prb,superscriptaddress,fleqn]{revtex4-2}
\usepackage[utf8]{inputenc}
\usepackage[english]{babel}
\usepackage{amsfonts}
\usepackage{graphicx}
\usepackage{xcolor}
\usepackage{hyperref}
\usepackage{bbold}
\usepackage{bm}

\newcommand{\br}[1]{\left( #1 \right)}
\newcommand{\bra}[1]{\left< #1 \right|}
\newcommand{\ket}[1]{\left| #1 \right>}
\newcommand{\ofk}{(\vec{k})}
\newcommand{\abs}[1]{\left| #1 \right|}
\newcommand{\dif}{\text{d}}
\newcommand{\kp}{{\vec{k}_\parallel}}
\renewcommand\vec{\bm}

\graphicspath{{figures/}}

\begin{document} 


\title{Berry curvature associated to Fermi arcs in continuum and lattice Weyl  systems}

\author{Dennis Wawrzik}
\affiliation{Institute for Theoretical Solid State Physics, IFW Dresden, Helmholtzstr. 20, 01069 Dresden, Germany}
\author{Jeroen van den Brink}
\affiliation{Institute for Theoretical Solid State Physics, IFW Dresden, Helmholtzstr. 20, 01069 Dresden, Germany}
\affiliation{Institute for Theoretical Physics and W\"urzburg-Dresden Cluster of Excellence ct.qmat, TU Dresden, 01069 Dresden, Germany}

\date{\today}

\begin{abstract}
Recently it has been discovered that in Weyl semimetals 
the {\it surface} state Berry curvature can diverge in certain regions of momentum. This occurs in a continuum description of tilted Weyl cones, which for a slab geometry results in the Berry curvature dipole associated to the surface Fermi arcs growing linearly with slab thickness. 
Here we investigate analytically incarnations of lattice Weyl semimetals and demonstrate this diverging surface Berry curvature by solving for their surface states and connect these to their continuum descriptions. We show how the shape of the Fermi arc and the Berry curvature hot-line is determined and confirm the $1/k^2$ divergence of the Berry curvature at the end of the Fermi arc as well as the finite size effects for the Berry curvature and its dipole, using finite slab calculations and surface Green's function methods. We further establish that apart from affecting the second order, non-linear Hall effect, the divergent Berry curvature has a strong impact on other transport phenomena as the Magnus-Hall effect and the non-linear chiral anomaly.
\end{abstract}

\maketitle

\section{Introduction}

The Berry curvature associated to electronic band-like states is a fundamental quantity that gives rise to an anomalous electronic velocity and interesting (non-linear) electronic, heat and spin-transport properties\cite{PhysRev.95.1154,PhysRevLett.93.046602,PhysRevLett.97.026603,RevModPhys.82.1959,PhysRevLett.105.026805,PhysRevLett.115.216806,zhang2018electrically,PhysRevB.98.121109,PhysRevB.99.201410,zhang2021different}. Recently it has been shown that electronic states at surface can have BC properties very different from bulk. For instance bulk BC-free systems exhibit a finite surface BC at certain surfaces, in particular ones of lower symmetry\cite{Wawrzik2022SurfaceIE} and surface Fermi arcs in Weyl semimetals have been shown to generically be accompanied by a surface Berry curvature that may diverge close to a hot line in the surface Brillouin zone connecting the projection of Weyl nodes with opposite chirality\cite{PhysRevLett.127.056601}. In Weyl semi-metals such surface Berry curvature appears whenever a bulk node has a velocity tilt toward the surface of interest\cite{PhysRevLett.127.056601,Ovalle2022}.

Here we consider such Weyl semimetals and the BC associated to their surface Fermi arcs in detail, in particular in the intrinsically regularized lattice setting with an even number of Weyl nodes. We first consider analytically a semi-infinite system --with a single surface-- both on the lattice and in the continuum. We show that the BC of the Fermi arc diverges quadratically along a hot-line in reciprocal space in both cases. We show how the free parameter that appears in the boundary condition of the self-adjoint extension in the continuum setting is regularized and becomes fixed by the lattice. Based on this insight we provide a time reversal invariant lattice Hamiltonian that allows tuning of this parameter, which strongly affects the Fermi arcs and their connectivity.

Subsequently we consider a finite size slab of size $L$ with both a bottom and top surface and show how the hybridization of the top and bottom Fermi arc and the size dependence of the BC dipole $D \propto L$ on the lattice confirm continuum expectations. Furthermore, we investigate the surface BC using the surface Green's function of the lattice model. Finally we consider Berry curvature mediated transport phenomena and show that the divergent surface BC has a significant influence on several phenomena like the Magnus Hall effect, second order Hall/Nernst effects, and non-linear chiral anomaly.

\section{Berry curvature of surface states}
To illustrate how the Berry curvature of surface states generally differs from that of bulk states, it is instructive to consider first a generic {\it Ansatz} for the wave function of a surface state $\Psi \br{z,\kp }$ where $z<0$ denotes the real space coordinate along the surface normal and $\kp$ the momentum parallel to the surface\cite{Wawrzik2022SurfaceIE}:
\begin{align}
\label{Eq:SurfaceState}
  \Psi \br{z,\kp } = c(\kp) ~f(z) e^{\lambda(\kp) z} \psi(\kp)
\end{align}
The surface state decays exponentially into the bulk where $\lambda(\kp)$ is complex with Re$(\lambda)>0$ and might be modulated by a general function $f(z)$. Without loss of generality we assume that the spinor part of the surface state $\psi^\dagger\psi=1$ is normalized which leads to the normalization constant
\begin{align}
  c^{-2} = \int_{-\infty}^0 \textup{d}z~|f|^2 e^{2\text{Re}(\lambda) z}.
\end{align}
The Berry connection $\vec{A}=-i\int_{-\infty}^0 ~\Psi^\dagger\nabla \Psi$ of the surface state becomes
\begin{align}
\label{Eq:Asurf}
  \vec{A} &= \vec{A}^\psi -i \left[ \frac{\nabla c}{c} + \bar{z}\nabla \lambda \right]
\end{align}
where $\vec{A}^\psi = -i\psi^\dagger\nabla\psi$ is the expected Berry connection of the spinor $\psi(\kp)$ and $\bar{z}(\kp)$ the expectation value of $z$ at a given momentum $\kp$:
\begin{align}
\label{Eq:barz}
  \bar{z}(\kp) &= c^2 \int_{-\infty}^0 \textup{d}z~z~|f|^2 e^{2\text{Re}(\lambda) z}
\end{align}
While $\nabla c/c=\nabla\log (c)$ is a pure gradient and vanishes in the Berry connection $\Omega=\nabla\times\vec{A}$, the last term in Eq.\ (\ref{Eq:Asurf}) depending on $\lambda$ does not. This means, that the BC $\Omega=\Omega^\psi+\Omega^\lambda$ of a surface state does not only depend on the BC of the spinor $\psi$ but has, in contrast to bulk states, a contribution $\Omega^\lambda$ originating from the momentum dependence of the localization of the surface state. More precisely, we have
\begin{align}
  \Omega^\lambda = 2\text{Re}(\nabla \lambda) \times \text{Im}(\nabla \lambda)~(\bar{z^2}-\bar{z}^2)
\end{align}
where $\bar{z^2}$ is defined analogous to Eq.\ (\ref{Eq:barz}). The BC is proportional to the variance $\bar{z^2}-\bar{z}^2$ and at the transition from a surface to a bulk state the state will spread through the whole crystal, i.e.\ Re$(\lambda)\rightarrow 0$, and the BC diverges. This divergence can be seen more clearly in the example $f(z) = z^n$ where  $n\in\mathbb{N}_0$. Here the normalization constant is
\begin{align}
  c = 2^n \sqrt{\frac{\br{\text{Re}(\lambda)}^{2n+1}}{n ~(2n-1)!}}
\end{align}
and $c = \sqrt{2\text{Re}(\lambda)}$ for $n=0$. From this we can calculate the Berry connection $A^\lambda = i (2n+1) \nabla \lambda/ \text{Re}(\lambda)$ and the corresponding BC is
\begin{align}
  \Omega^\lambda = \br{n+\frac{1}{2}}~\frac{\text{Re}(\nabla \lambda)\times\text{Im}(\nabla \lambda)}{(\text{Re}(\lambda))^2}.
\end{align}
From this expression it is clear that the BC associated to the surface state may diverges at the line for which Re$(\lambda)=0$. As we will see later on, precisely this is case for Fermi arcs in tilted Weyl semimetals with $n=0$ and $\lambda\propto \kp$: the end points of a Fermi arc always hit the BC hot-line such that the BC diverges quadratically $\Omega\propto \kp^{-2}$.

\section{Surfaces of semi-infinite Weyl systems}
The linear dispersion around a Weyl node has the remarkable consequence that on the surface a  topological protected Fermi arc is induced, which connects the surface projections of two Weyl nodes of opposite chirality\cite{Wan2011,Armitage2018}. Before investigating these surface states and their Berry curvature in a lattice setting, we will first consider the simpler case of a continuum description for a single tilted Weyl node and its associated surface Berry curvature. 

\subsection{Single tilted Weyl cone in continuum}
Even if in condensed matter settings Weyl cones come in pairs of opposite chirality\cite{Nielsen1981a,Nielsen1981b}, one may consider a low energy model of a single general, tilted Weyl cone with chirality $\chi=\pm1$ given by the Hamiltonian
\begin{align}
    H = -i\vec{\nabla}_{\vec{r}} \cdot \br{\chi V \pmb \sigma + \vec{u} \sigma_0} + \mu \sigma_0
    \label{eq:H_kdotp}
\end{align}
where $\sigma_i$ are the Pauli matrices, $\vec{u}$ the tilting vector, $\mu$ the energy of the node, and $V=V^T$ a symmetric matrix with $\det(V) > 0$ allowing for an anisotropic velocity of the Weyl cone. We restrict our model to a type-I Weyl cone where $V^2 - \vec{u}\vec{u}^T$ is positive definite. Furthermore, it is convenient to simplify the model by setting $V$ to the identity matrix, $\mu=0$, and $\vec{u}= (0,0,u_z)$ and restore everything later on. Without loss of generality we choose the surface to be at $z=0$ and restrict the Hamiltonian in Eq.\ (\ref{eq:H_kdotp}) to the lower half space $z<0$. Therefore, we split every vector in a parallel and perpendicular part with respect to the surface, i.e.\ $\vec{u} = \vec{u}_\parallel + u_z\hat{z}$.

In this setup, hermiticity $\left< \psi,H \psi\right> = \left<H \psi, \psi\right>$ requires any surface state $\psi$ to have $\psi^\dagger \br{\chi \sigma_z + u_z}\psi |_{z=0} = 0$ at the surface, i.e.\ the $z$-component of the pseudo-spin expectation value $s_z = -\chi u_z$ compensates for the tilt such that there is no net current into the vacuum. Since there is no further constraint on $\vec{s}_\parallel$ besides $|\vec{s}|=1$ it can be parametrized by a real number $\alpha$ (see also Section \ref{Sec:selfadjoint}) as follows:
\begin{align}
\label{eq:s}
	\vec{s} &= \left< \psi_\alpha | \pmb \sigma | \psi_\alpha \right> \nonumber\\
	        &= \left(\sqrt{1-u_z^2} \cos (\alpha), \sqrt{1-u_z^2} \sin(\alpha), - \chi u_z \right)
\end{align}
With this we are able to make an ansatz for a wave function $\Psi$ that decays exponentially away from the surface with inverse decay length $\lambda$ similar to Eq.\ (\ref{Eq:SurfaceState}):
\begin{align}
	\label{eq:ansatz}
	\Psi\br{\vec{k_\parallel},z} = c\br{\kp} e^{i\kp\cdot\vec{r}_\parallel + \lambda\br{\kp} z} \psi_\alpha
\end{align}
Here, $\psi_\alpha$ is the spinor fulfilling Eq.\ (\ref{eq:s}) so that the boundary condition is always satisfied. The normalization constant $c$ is determined by $1 = \int_{-\infty}^0 \dif z \abs{\Psi}^2$ which gives $c^2 = 2$Re$(\lambda)$ and Re$(\lambda) > 0$. Before we can get the actual value of $\lambda$ we first have to calculate the energy of the surface state
\begin{align}
	\label{eq:energy_wo_tilt}
	E(\kp)  &= \left< \Psi |H| \Psi \right> \nonumber\\
	        &= \br{\vec{k}_\parallel-i\lambda\hat{z}} \cdot \br{\chi\vec{s}+u_z\hat{z}} \nonumber\\
	        &= \chi\kp \cdot \vec{s}_\parallel
\end{align}
where the part proportional to $-i\lambda$ vanishes because of the boundary condition $s_z=-\chi u_z$.
For $E=0$ this energy dispersion gives rise to a Fermi arc perpendicular to $\vec{s}_\parallel$, the exact direction will be fixed by the requirement Re$(\lambda) > 0$. In order to find a solution for $\lambda(\kp)$ it is convenient to multiply the Hamiltonian by $\sigma_z$, i.e.\ the equation $-\chi u_zE(\kp) = \bra{\Psi} H \sigma_z\ket{\Psi}$ yields:
\begin{align}
\label{eq:lambda}
	\lambda = -\frac{\kp\cdot \br{\hat{z}\times\vec{s} + i \chi u_z \vec{s}}}{1-u_z^2}
\end{align}
Now we have the full wave function and can calculate the BC $\Omega_z(\kp)= \pmb \nabla_{\kp}\times \vec{A}_\parallel$ of the surface state with Berry connection
\begin{align}
  \vec{A}_\parallel (\kp) = -i \int_{-\infty}^0 \dif z~ \Psi^\dagger\br{\vec{k_\parallel},z} \pmb\nabla_{\kp} \Psi\br{\vec{k_\parallel},z}
\end{align}
to be:
\begin{align}
\label{eq:BC}
  \Omega_z(\kp) &= \frac{\text{Im}(\vec{\nabla}\lambda)\times \text{Re}(\vec{\nabla}\lambda)}{2\text{Re}^2(\lambda)}\\
                &= -\frac{\chi u_z \br{1-u_z^2}}{2\br{\kp\cdot\br{\hat{z}\times\vec{s}}}^2}\hat{z}
\end{align}
The resulting BC is proportional to the tilt towards the surface and diverges at the BC hot-line where $\kp \propto \vec{s}_\parallel$.

Having the results of the simplified model, we can add the terms we skipped. The terms depending on $\vec{u}_\parallel$ and $\mu$ are proportional to $\sigma_0$ and thus do not affect $\Psi$ and $\lambda$ but directly add to the energy $E(\kp)$. The easiest way to include the velocity matrix is to shift $V$ in the Hamiltonian in Eq.\ (\ref{eq:H_kdotp}) to the derivative which is basically done by substituting $(\kp,\hat{z})\rightarrow (V\kp,V\hat{z})$ and $\vec{u}\rightarrow V^{-1}\vec{u}$. The boundary condition then becomes $\vec{s}\cdot(V\hat{z})=-\chi u_z$ and
\begin{align}
  E &=  \chi(V\kp) \cdot \vec{s}_\parallel + \kp\cdot\vec{u}_\parallel + \mu \\
  \lambda &= (V\kp) \cdot \frac{\vec{s}\times V\hat{z} - i \br{V\hat{z} + \chi u_z \vec{s}}}{|V\hat{z}|^2-u_z^2}
\end{align}
yielding
\begin{align}
\vec{\Omega} = -\frac{ \br{\left| V\hat{z} \right|^2 - u_z^2} \br{V^{-1}\vec{s}\cdot \hat{z}}}{2 \det(V) \br{\kp \cdot \br{V^{-1}\vec{s} \times \hat{z}}}^2} \hat{z}.
\end{align}
For anisotropic Weyl cones it is possible that no principal axis of $V$ is pointing along $\hat{z}$ such that $(V\kp) \cdot (V\hat{z})$ is finite. Thus, we can have Im$(\vec{\nabla}\lambda) \ne 0$ even if $u_z=0$ and, according to Eq.\ (\ref{eq:BC}), find BC without any tilt.

\subsection{Lattice Weyl systems}
As states above, according to the Nielsen-Ninomiya theorem\cite{Nielsen1981a,Nielsen1981b} Weyl nodes in crystals always come in pairs with opposite chirality, which implies that our lattice model must have at least two nodes. Similar to Ref.\ \cite{PhysRevB.95.075133}, we may define a tilted Weyl cone pair Hamiltonian $H_\text{L}\ofk = H_\text{cone} + H_\text{tilt}+\mu\sigma_0$ on a cubic lattice with lattice constant $a=1$ as follows:
\begin{widetext}
    \begin{align}
    \label{eq:H_pair}
	    H_\text{cone} =& \chi\sin(k_y)\br{V\vec{\sigma}}_y + \chi\sin(k_z)\br{V\vec{\sigma}}_z + \left[\chi \br{\cot(k_0) - \frac{\cos(k_x)}{\sin(k_0)}} + \Delta \br{2 - \cos(k_y) - \cos(k_z)}\right] \br{V\vec{\sigma}}_x\\
	    H_\text{tilt} =& \left[u_y\sin(k_y) + u_z\sin(k_z) + \vphantom{\frac{1}{1}} u_x\br{\cot(k_0) - \frac{\cos(k_x)}{\sin(k_0)}} \right] \sigma_0
    \end{align}
\end{widetext}
This model has Weyl cones at $(\pm k_0,0,0)$ with chirality $\pm \chi$. The cone at $k_x=+k_0$ has exactly the dispersion described by the $k \cdot p$ model in Eq.\ (\ref{eq:H_kdotp}) while the other one is related by a mirror symmetry $M_x$. Again, for simplicity we take $V = \mathbb{1}$, $k_0 = \pi/2$, $\chi=\Delta=1$, and $u_x=u_y=0$. It is worth to notice that, compared to the continuum $k \cdot p$ model, we have an extra term $\Delta\br{2-\cos (k_y) - \cos (k_z)}\sigma_x$. It vanishes at the Weyl nodes but is needed to open a gap of size $4\Delta$ at $k_y=\pi$ and $k_z=\pi$ planes. As in the previous section, we construct a surface by constraining the Hamiltonian to $n \leq 0$ where $n$ labels the lattice sites in $z$-direction. In the vacuum $n>0$ no hoppings are allowed. We can keep the Fourier transform in $\kp$-space parallel to the surface but have to use the real space hoppings perpendicular to it:
  \begin{eqnarray}
	H(\kp) &=& \sum_{n=-\infty}^{0} \vec{c}^\dagger_n h_\parallel(\kp) \vec{c}^{\phantom{\dagger}}_n  \nonumber \\
	&+& \sum_{n=-\infty}^{0} \br{\vec{c}^\dagger_n t \vec{c}^{\phantom{\dagger}}_{n+1} + \vec{c}^\dagger_n t^\dagger \vec{c}^{\phantom{\dagger}}_{n-1}}
  \end{eqnarray}
where
\begin{align}
	h_\parallel(\kp) &= \br{2 - \cos(k_x) - \cos(k_y)}\sigma_x + \sin(k_y) \sigma_y\\
	t &= -\frac{1}{2}\br{\sigma_x + i\br{\sigma_z+u_z\sigma_0}}
\end{align}
and $\vec{c}^\dagger_n$ is the creation operator for a particle in the $n$th layer with momentum $\kp$.

Now the solution of the surface state $\vec{\Psi}(\kp)$ in this semi-infinite slab lattice model can be derived in a similar manner as of the continuum model. For any state  $\vec{\psi}=\br{\dots ,\psi_{n-1},\psi_n,\psi_{n+1},\dots}$ a hermitian Hamiltonian has to fulfil
\begin{align}
	0 &= \langle \vec{\psi} | H \vec{\psi} \rangle - \langle H \vec{\psi} | \vec{\psi} \rangle\\
	&= 2 \,\text{Im}\br{\psi_{0}^\dagger t \psi^{\phantom{\dagger}}_1} 
\end{align}
where all terms cancel except the ones at the boundary. If we assume that, as in the continuum case, the normalized 2-spinor $\psi_n\equiv\psi$ is independent of $z$ or rather $n$ we arrive at the known boundary condition
\begin{align*}
	\psi^\dagger \br{t-t^\dagger} \psi=0\quad\Leftrightarrow\quad s_z=\psi^\dagger\sigma_z\psi=-u_z.
\end{align*}
In addition, there should be effectively no current into vacuum, i.e.\ $\psi^\dagger t \psi=0$. This way, the additional $\Delta$ term in Eq.\ (\ref{eq:H_pair}) yields $s_x=0$ and the previously free parameter $\alpha$ is fixed to be $\alpha=\pm\pi/2$ depending on the sign of $\Delta$. This is expected because the lattice Hamiltonian is a bounded operator, i.e.\ there exists an $E_0 >0$ such that $|H\Psi|\le E_{0}|\Psi|$ for all $\Psi$, and therefore does not need an extension.

\begin{figure}
  \centering
  \includegraphics[width=\columnwidth]{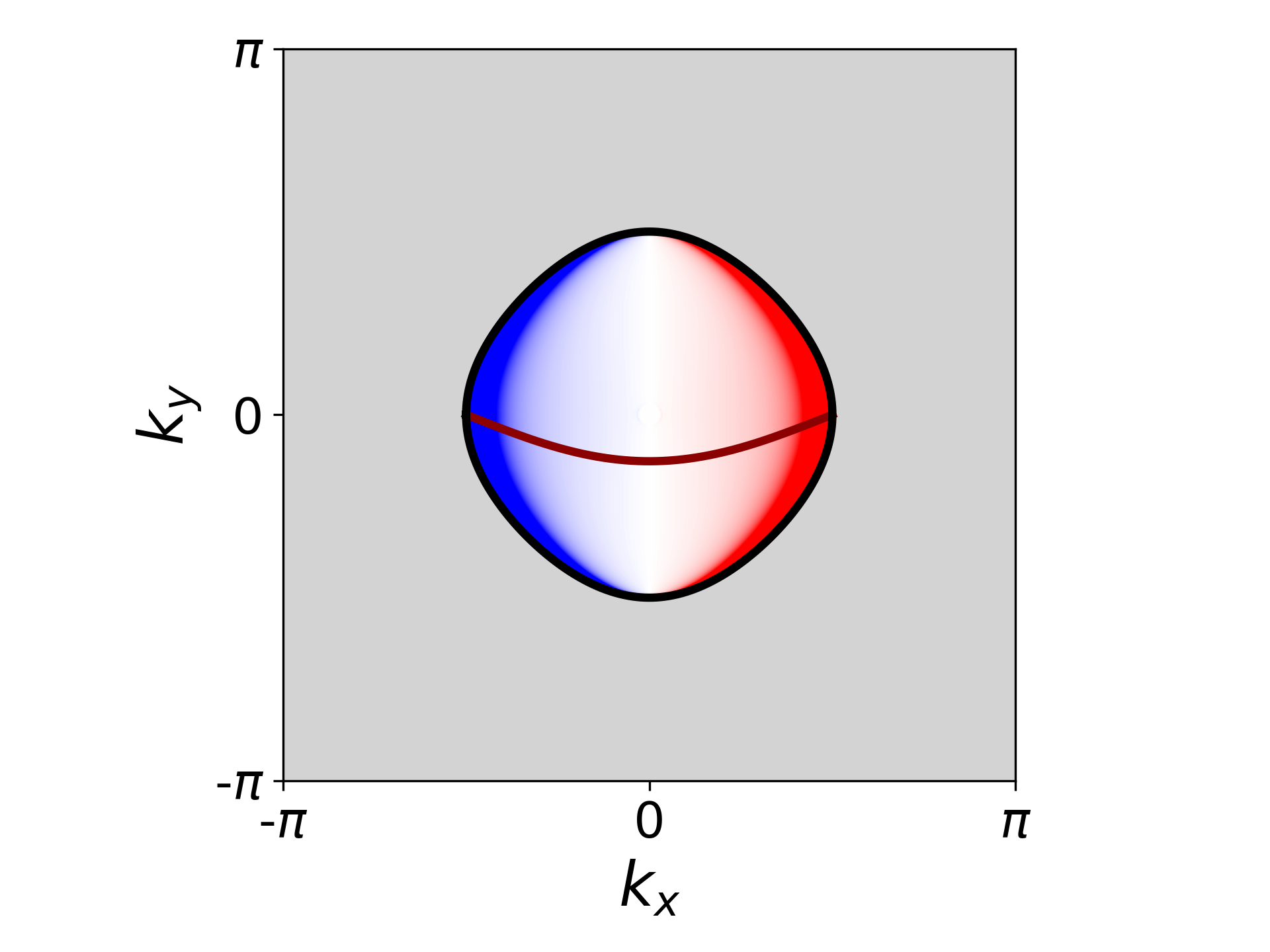}
  \caption{Berry curvature, Fermi arc (red line), and hot-line (black line) of 
   the tilted Weyl cone pair Hamiltonian $H_\text{L}\ofk = H_\text{cone} + H_\text{tilt}$ of Eq.\ (\ref{eq:H_pair}) on a cubic lattice. The parameters are $V = \mathbb{1}$, $k_0 = \pi/2$, $\chi=\Delta=1$ and the tilt is $\vec{u} = (0, 1/4,1/2)$. The gray area denotes delocalized bulk states.}
  \label{fig:FA_HL_LM}
\end{figure}

As in Eq.\ (\ref{eq:ansatz}) we make an ansatz for the surface wave function $\vec{\Psi} = \br{\dots ,\Psi_{-2},\Psi_{-1},\Psi_0, 0,\dots}$ which decays exponentially into the bulk and is proportional to the above $\psi$:
\begin{align}
	\Psi_n(\kp) = c(\kp) ~r^{-n}(\kp)\,\psi
\end{align}
with $n\leq 0$ and $|r|<1$. We find $c^2 = 1-|r|^2$ and
\begin{align}
	\br{H\vec{\Psi}}_n &= h_\parallel \Psi_n + t\Psi_{n+1} + t^\dagger\Psi_{n-1}\\
	&= \br{h_\parallel + r t+r^{-1}t^\dagger} \Psi_n
\end{align}
As in the continuum case we can use Schr\"odinger's equation and $-u_zE = \Psi^\dagger_n H_{\text{eff}}\sigma_z\Psi^{\phantom{\dagger}}_n$ to solve for energy $E(\kp)$ and the complex number $r(\kp)$. We get
\begin{align}
	E(\kp) &= (-\sqrt{1-u_z^2} + u_y) ~\sin(k_y) + u_x \sin(k_x)\\
	r(\kp) &= \frac{z+\sqrt{z^2-u_z^2}}{1+\sqrt{1-u_z^2}}
	\label{eq:r}
\end{align}
where we restored $u_x$ and $u_y$, and
\begin{align}
	z = 2-\cos(k_x)-\cos(k_y)+i u_z \sin(k_y).
\end{align}
At the Weyl nodes $k_x=\pm \pi/2$ and $k_y=0$ we obtain $r=z=1$, i.e. the state becomes a surface state as expected. Without a tilt Eq.\ (\ref{eq:r}) simplifies to $r=z$ everywhere and becomes a real number. The Berry curvature is given by
\begin{align}
\label{eq:BC_L}
	\Omega_z = 2\,\frac{\text{Im}(\vec{\nabla}r)\times \text{Re}(\vec{\nabla} r)}{\br{1-|r|^2}^2}
\end{align}
and diverges at the hot-line where $|r| = 1$ as shown in Fig.\ \ref{fig:FA_HL_LM}.

If the lattice becomes denser and denser we can connect the results from the lattice model to the continuum limit. Therefore, we introduce a small lattice constant $a$, i.e.\ using $z = na$ and $\sin(ak_i)$ or $\cos(ak_i)$ where we let $a\rightarrow 0$. In this manner we find on the one hand from Eq.\ (\ref{eq:r}) close to the Weyl point at $k_x=\pi/2$
\begin{align}
  r \approx 1 - a\frac{k_x +iu_zk_y}{\sqrt{1-u_z^2}}
\end{align}
and on the other hand $\exp(\lambda z) \approx (1-a\lambda)^{-n}$
which yields $r \approx 1-a\lambda$ consistently with Eq.\ (\ref{eq:lambda}). This also gives $c^2=1-|r|^2\approx 2a\text{Re} (\lambda)$ and $\nabla r \approx -a\nabla \lambda$. It turns out that we get the same BC as in Eq.\ (\ref{eq:BC_L}) for the lattice model with finite $a$ if we use the continuum BC in Eq.\ (\ref{eq:BC}) and these approximations, even though both are valid close to the Weyl nodes and for $a\rightarrow 0$ only.

\subsection{Self-adjoint extensions and parameter $\alpha$}
\label{Sec:selfadjoint}

Symmetric, unbound operators like the differential operator $P = -i\partial_x$ in systems with boundaries are in general not self-adjoint\cite{doi:10.1119/1.1328351,10.1093/ptep/ptx053,PhysRevB.97.075132}. Nevertheless, they can have self-adjoint extensions which usually come in form of boundary conditions. For example, if we consider wave functions $\psi$ and $\phi$ constraint to the interval $[0,1]$ we find by partial integration
\begin{align}
\label{eq:P}
  \left<\phi| P \psi\right> - \left<P\phi|\psi\right> = -i \left[\phi^\dagger(1) \psi(1) - \phi^\dagger(0) \psi(0)\right].
\end{align}
Even though $P^\dagger$ (acting on $\phi$) formally looks the same as $P$ (acting on $\psi$) they do not act on the same space of functions: for a particle in a box we usually take the boundary conditions $\psi(0)=\psi(1)=0$ which always yield zero for the right hand side of Eq.\ (\ref{eq:P}) but do not give any restrictions on $\phi$. In order to make $P$ self-adjoint we have to find a boundary condition for which the boundary term in Eq.\ (\ref{eq:P}) vanishes and which is the same for both $P$ and $P^\dagger$. The easiest of these so-called self-adjoint extensions is $P_0$ with periodic boundary conditions $\psi(0)=\psi(1)$. Indeed, there are infinitely many possible extensions parametrized by a phase $\theta$ as $P_\theta$ with $\psi(0)=e^{i\theta}\psi(1)$. According to von Neumann's theorem\cite{v1930allgemeine}, if an operator has a self-adjoint extension it is either unique or there are infinitely many extensions parametrized by a unitary matrix $\mathbb{U}(n)$. 

As a consequence, we find a $\mathbb{U}(1)$ phase $e^{i\alpha}$ for the tilted Weyl cone model in Eq.\ (\ref{eq:H_kdotp}) at the boundary $z=0$
\begin{align}
  \psi_\downarrow=e^{i\alpha}\sqrt{\frac{1+\chi u_z}{1-\chi u_z}}~\psi_\uparrow
\end{align}
which is equivalent to the condition $\bra{\psi} \pmb \sigma \ket{\psi} = \vec{s}(\alpha)$ in Eq.\ (\ref{eq:s}) yielding
\begin{align}
  \psi(\alpha) = \frac{1}{\sqrt{2}}
	\begin{pmatrix}
		e^{-i\alpha/2} \sqrt{1-\chi u_z} \\
		e^{+i\alpha/2} \sqrt{1+\chi u_z}
	\end{pmatrix}.
\end{align}
Note that the choice of $\alpha$ has direct physical consequences since it defines the direction of the Fermi arc in momentum space.

If a hermitian Hamiltonian is defined on the whole space it always is self-adjoint since the wave functions vanish at $\pm \infty$ and there is no boundary term anymore. This means if we define the insulating phase at $z>0$ it will fix a specific value of $\alpha$. As example, we take the Hamiltonian  in Eqs.\ (\ref{eq:H_kdotp}) and substitute $k_y$ with
\begin{align}
  \tilde{k}_y(z) = 
  \begin{cases}
    \frac{+k_0^2-k_y^2}{2k_0}, & z\leq 0\\
    \frac{-\Delta^2-k_y^2}{2\Delta}, & z>0
  \end{cases}.
\end{align}
This yields two Weyl cones at $\vec{k}=(0,\pm k_0,0)$ with opposite chirality for $z<0$ and an insulator with gap $2\Delta$ for $z>0$, see SM of Ref.\ \cite{PhysRevLett.127.056601}. Again, the wave function decays exponentially $\psi \propto e^{\lambda(\kp) z}$ away from the surface $z=0$, i.e.\ Re$(\lambda z)<0$. From Eqs. (\ref{eq:s}) and (\ref{eq:lambda}) we find
\begin{align}
\text{Re}(\lambda) \propto -\sin(\alpha) k_x  + \cos(\alpha) \tilde{k}_y(z).
\end{align}
Thus, on the insulating site Re$(\lambda z)<0$ is only satisfied for all $\kp$ if $\sin(\alpha)=0$ and $\cos(\alpha)>0$, i.e. $\alpha = 0$. For the Weyl semimetal side the surface state has to be located at $|k_y|<k_0$ and the Fermi arc becomes a straight line between the Weyl nodes. Even though this holds for the perfect vacuum $\Delta\rightarrow\infty$, too, other conventions for the insulator side will yield different values for $\alpha$. 

\begin{figure}
  \begin{minipage}{0.238\textwidth}
    \includegraphics[width=\textwidth]{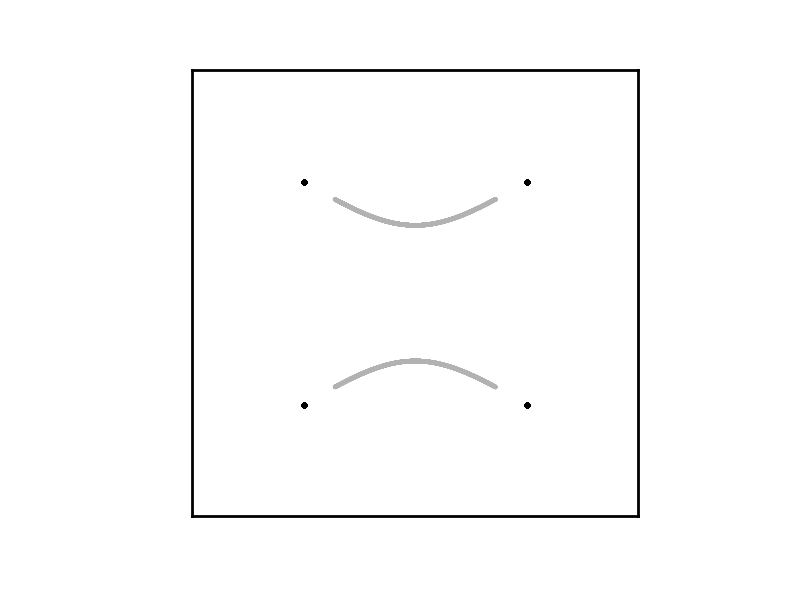}
    $\gamma=\pi/6$
  \end{minipage}
  \begin{minipage}{0.238\textwidth}
    \includegraphics[width=\textwidth]{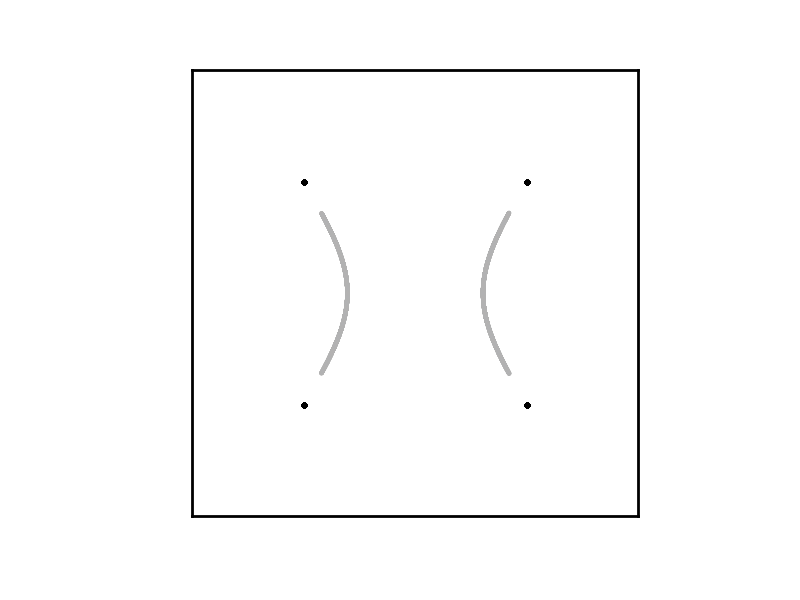}
    $\gamma=\pi/3$
  \end{minipage}
  \begin{minipage}{0.238\textwidth}
    \includegraphics[width=\textwidth]{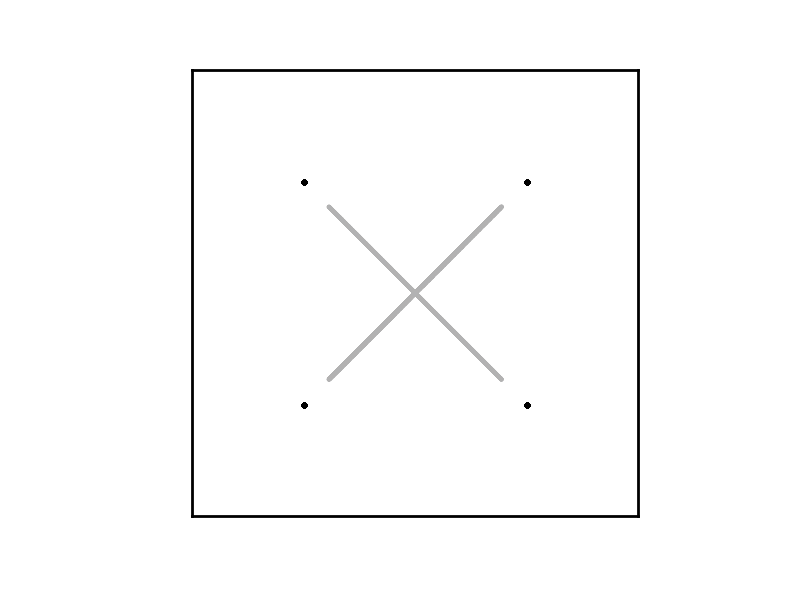}
		$\gamma=\pi/4,~\mu=0$
  \end{minipage}
  \begin{minipage}{0.238\textwidth}
    \includegraphics[width=\textwidth]{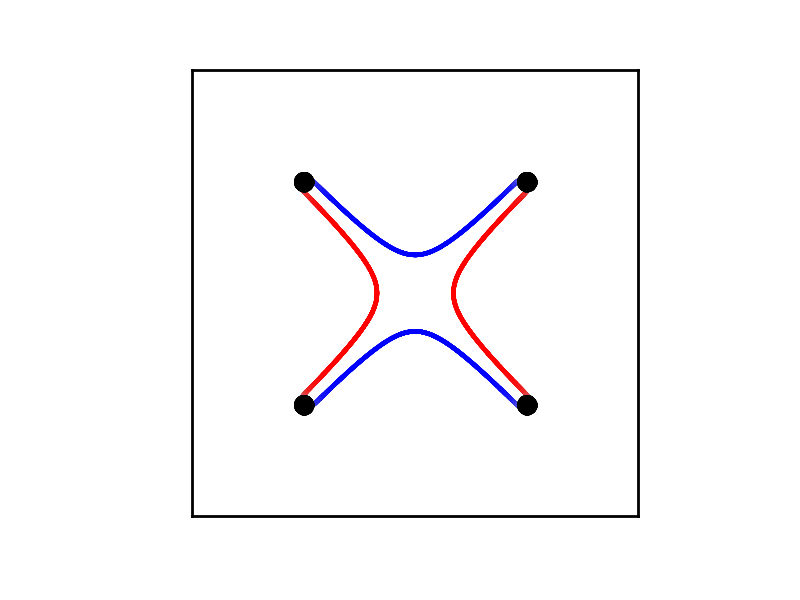}
    $\gamma=\pi/4,~\mu=0.1$
  \end{minipage}
  \caption{Fermi arcs of the lattice model with 4 Weyl nodes on surface (001) for different values of $\gamma$ and $\mu$. The projections of the bulk Weyl cones are marked in black. A red (blue) arc is located on the top (bottom) surface while for a gray both arcs are on top of each other.}
  \label{fig:Weyl4_alpha}
\end{figure}

As shown in the previous section, this parameter will be fixed in a lattice model, too. The model in Eq.\ (\ref{eq:H_pair}) only allows for two different values $\alpha=\pm\pi/2$ by tuning $\Delta$ that opens a gap on the boundary of the Brillouin zone. However, introducing the time reversal invariant~\cite{McCormick2017} lattice model
\begin{align}
  \label{eq:H_TR}
  H_\text{TR} = -&\cos(k_x)\sigma_x - \cos(k_y)\sigma_y + \sin(k_z)\sigma_z + \nonumber \\
		       & \Delta (1-\cos(k_z))\left[\cos(\gamma) \sigma_x + \sin(\gamma) \sigma_y\right] 
\end{align}
with four Weyl nodes at $k=(\pm \frac{\pi}{2},\pm \frac{\pi}{2},0)$ we only have to gap the cones at $k_z=\pi$. Now we have the choice to use the $\Delta$ term with either $\sigma_x$ or $\sigma_y$ allowing to interpolate between these two via $\gamma$. On the (001) surface the energy dispersion reads
\begin{align}
	E = \pm\br{\cos(k_x)\sin(\gamma) - \cos(k_y)\cos(\gamma)}
\end{align}
and one finds $\alpha = \gamma$. As shown in Fig.\ \ref{fig:Weyl4_alpha}, variation of $\gamma$ changes the shape of the Fermi arcs in such a way that one can have different connectivity between the Weyl nodes. Furthermore, we can introduce a tilt $H_{tilt} = -\br{u_x\cos(k_x)+u_y\cos(k_y)}\sigma_0$ whereby a tilt in $z$-direction is forbidden if we want to keep (spinless) time reversal $\Theta=\sigma_x K$. In this case it is even possible to have different connectivity on top and bottom surfaces as illustrated in Fig.\ \ref{fig:Weyl4_alpha}.

\section{Finite-size Weyl slabs}
In order to study finite size effects in Weyl semimetals we take the same Hamiltonian and ansatz as in the semi-infinite case from Eqs.\ (\ref{eq:H_kdotp}) and (\ref{eq:ansatz}) but restrict them to a finite interval $|z| < L/2$. This time it is convenient to solve Schr\"odinger's equation for $\lambda$ first which results in
\begin{align}
	\lambda_s &= \frac{s\sqrt{k_\parallel^2-\epsilon^2} - i u_z \epsilon}{\sqrt{1-u_z^2}}\\
	\psi_s &= \frac{1}{\sqrt{2}}
	\begin{pmatrix}
		\frac{\epsilon - i s \sqrt{k_\parallel^2-\epsilon^2}}{k_x + i k_y} \sqrt{1-\chi u_z} \\
		\sqrt{1+\chi u_z}
	\end{pmatrix}
\end{align}
with $\epsilon(\kp)$ defined via the energy $E(\kp)$ of the surface state as $\chi\sqrt{1-u_z^2}\epsilon(\kp) = E(\kp) - \kp\cdot\vec{u}$. Now we have to consider both solutions with $\text{Re}(s\lambda)>0$ where $s=+1$ ($s=-1$) labels the wave function localized on the top (bottom) surface. The full wave function is a superposition of both solutions of $\lambda$:
\begin{align}
	\Psi\br{\kp,z} = \frac{e^{i\kp\cdot\vec{r}_\parallel}}{\sqrt{2}}\sum_{s=\pm 1} s c_s e^{\lambda_s z} \psi_s
\end{align}
For the finite case the boundary condition $\left<\psi|H\psi\right>-\left<H\psi|\psi\right>=0$ changes to
\begin{align}
\label{eq:psi_L}
	\psi_+^\dagger \br{\chi \sigma_z + u_z}\psi_+ = \psi_-^\dagger \br{\chi \sigma_z + u_z}\psi_-
\end{align}
where $\psi_\pm = \psi (\pm L/2)$ is the wave function at the boundary. The solution closest to the semi-finite case for this equation is that both sides of Eq.\ (\ref{eq:psi_L}) vanish and $\psi_\pm$ is determined by Eq.\ (\ref{eq:s}) where $\alpha_\pm$ can be chosen independently on both sides. The complete solution can be parametrized by a unitary matrix $U \in \mathbb{U}(2)$ as followes:
\begin{align}
  \begin{pmatrix}
    \sqrt{1-\chi u_z} \psi_{+\downarrow} \\
    \sqrt{1+\chi u_z} \psi_{-\uparrow} 
  \end{pmatrix}
  = U 
  \begin{pmatrix}
    \sqrt{1+\chi u_z} \psi_{+\uparrow} \\
    \sqrt{1-\chi u_z} \psi_{-\downarrow} 
  \end{pmatrix}
\end{align}
Even though $U$ can be parametrized by four real numbers we will concentrate on the above solution given by $U = \text{diag}(e^{i\alpha_+}, -e^{i\alpha_-})$ because the off-diagonal terms correspond to periodic boundary conditions and we will also find this choice of $U$ in the lattice model. This boundary condition yields an equation for the energy dispersion $\epsilon(\kp)$
  \begin{eqnarray}
  \label{eq:E_finite}
	\tanh\br{L \sqrt{\frac{k_\parallel^2-\epsilon^2}{1-u_z^2}}}  =  \nonumber \\  \frac{\sqrt{k_\parallel^2-\epsilon^2}\cos(\gamma)}{k_y\cos(\theta) - k_x \sin(\theta) - \epsilon \sin(\gamma)}
  \end{eqnarray}
  where $\alpha_\pm = \theta \pm \gamma$. For $\theta = 0$ normalization yields
  \begin{align}
	c_s^2 &=  c_0 \frac{\br{\br{k_+\sin(\gamma)-i\epsilon} \br{i\epsilon - s \sqrt{k_\parallel^2-\epsilon^2}}-k_x k_+}}{L \epsilon c_0^2 - \sqrt{1-u_z^2}\cos(\gamma)\br{k_y \epsilon - k_\parallel^2\sin(\gamma)}}, \nonumber \\
	c_0^2&= \br{\epsilon - k_y\sin(\gamma)}^2 - \br{k_x \cos(\gamma)}^2,
  \end{align}
where $k_+=k_x+ik_y$.
As shown in Ref.\ \cite{PhysRevLett.127.056601}, the surface states are given by the two solutions with $\epsilon^2 < k_\parallel^2$ and result in Fermi arcs that hybridize before they touch the Weyl node resulting in a finite BC dipole growing linearly in the slab size $L$. Note that the Weyl cone $\epsilon = \pm k_\parallel$ itself formally solves Eq.\ (\ref{eq:E_finite}) and results in a state $\psi(\kp)$ that satisfies the boundary condition Eq.\ (\ref{eq:psi_L}). Nevertheless, we find $\vec{s}_\parallel = \pm\sqrt{1-u^2}\kp/k_\parallel$ and therefore it is independent of $\alpha_\pm$. Thus, in a finite lattice model where $\alpha$ and the direction of  $\vec{s}_\parallel$ is fixed, the solution $\epsilon = \pm k_\parallel$ will not be found. In a lattice model the values $\alpha_\pm$ are fixed in the same way as in the semi-infinite case, i.e.\ by the $\Delta$ term in Eq.\ (\ref{eq:H_pair}) (Eq.\ (\ref{eq:H_TR})), and we find $\alpha_\pm = \pi/2$ ($\alpha_\pm=\gamma$) for the model with two (four) Weyl nodes. Therefore, the Fermi arcs on top and bottom surface have the same shape if $\vec{u}_\parallel=0$. 

\begin{figure}[h]
  \includegraphics[width=\columnwidth]{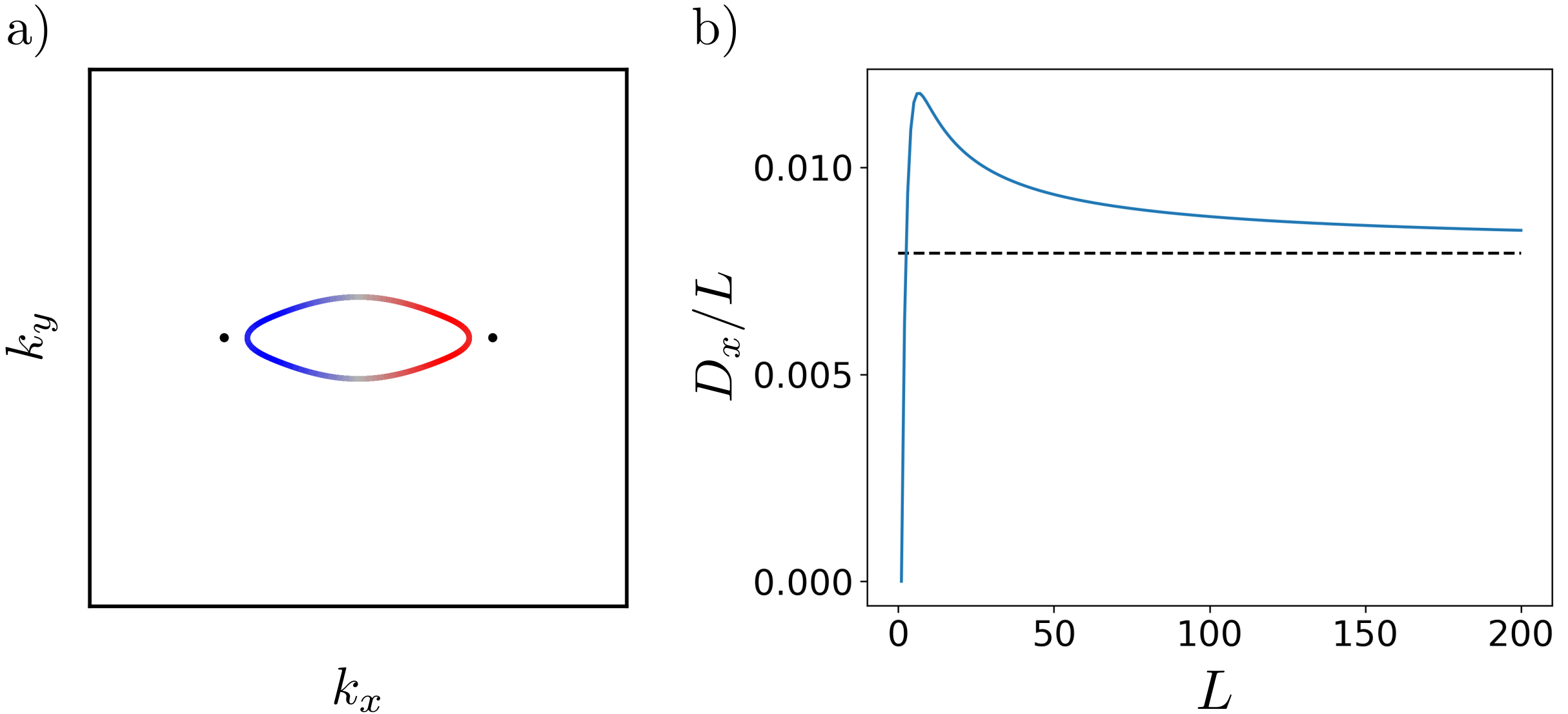}
  \caption{a) Fermi arc with Berry curvature of the lattice model with a single tilted Weyl pair on the $(001)$ surface. The tilt is $u_x = 1/2$ and $u_z=1/4$, with $V=\mathbb{1}$ and slab size $L=4$. The black dots mark the projections of the Weyl nodes. b) Surface Berry curvature dipole $\vec{D}$ in dependence of slab size $L$. The blue line corresponds to $D_x/L$ while the dashed line shows the behavior for large $L$. 
  }
  \label{fig:BC(D)_arc_L}
\end{figure}

The lattice model with two Weyl nodes in Eq.\ (\ref{eq:H_pair}) is well suited to numerically calculate the slab thickness dependence of the BC dipole. As illustrated in Fig.\ \ref{fig:BC(D)_arc_L} (a), the Fermi arcs of opposite sites hybridize before touching the Weyl points and have most BC closest to them because this is where the surface states are spread through the whole bulk and have a large variance. We can rewrite the BC dipole at $T=0$ as an integral along the Fermi arc
\begin{align}
  \vec{D} &= -\int \frac{\dif^2k}{(2\pi)^2}~ \vec{\nabla} f_0 ~\Omega(\kp) \\
          &= \int_{E(\kp)=0} \frac{\dif k}{(2\pi)^2}~ \Omega(\kp) \frac{\vec{v}_\parallel}{|\vec{v}_\parallel|}
\end{align}
where $\vec{v}_\parallel = \vec{\nabla} E$. As shown in Fig.\ \ref{fig:BC(D)_arc_L} (b), up to a correction of order $\mathcal{O}(a/L)$ due to the finite lattice constant $a$ this confirms the linear behavior of the BC dipole with $D_x \approx 0.008 L$. 

Another way to study the BC of Fermi arcs in a lattice model is to look at surface Green's functions. Starting from a Green's function (GF) $G(i\omega,{\bm k} ) = (i \omega - H_{\bm k})^{-1}$ the BC of all occupied states is given by \cite{ISHIKAWA1987523,PhysRevLett.113.136402}
\begin{equation}
  \Omega_c({\bm k})=\int_{-\infty}^{\infty}\frac{d\omega}{2 \pi} \text{tr}\br{g_c(i\omega,{\bm k})}
  \label{Eq:BC}
\end{equation} with
\begin{equation}
  g_c(i\omega,{\bm k} ) = \frac{\epsilon_{\mu \nu \rho c}}{3!}~G \br{\partial_\mu G^{-1}} G \br{\partial_\nu G^{-1}} G \br{\partial_\rho G^{-1}}
  \label{Eq:BC_Green}
\end{equation}
where $c \in \{x,y,z\}$ and $\mu,\nu,\rho \in \{\omega,x,y,z\}$. Following Ref.\ \cite{Wawrzik2022SurfaceIE}, we use the surface GF obtained via the iterative algorithm in \cite{san85} to calculate the surface BC. The algorithm efficiently provides the GF for the top $N$ layers of a semi-infinite slab geometry. The BC dipole in terms of GFs is given by
\begin{align}
	D^S_{a}(\mu) &= -\int \frac{\text{d}^2{\bm k}_\parallel}{(2\pi)^2} ~\partial_a f(\mu,\vec{k}_\parallel) ~\Omega^S_z({\bm k_\parallel}) \nonumber \\
              &= -\int \frac{\text{d}^2{\bm k}_\parallel}{(2\pi)^2} ~d^S_{a}(\mu,{\bm k}_\parallel)\\
	d^S_{a}(\mu,{\bm k}_\parallel) &=\int_{-\infty}^\infty \frac{\text{d} \omega}{2\pi}~\text{tr} \Big[ \partial_a \big(G^{S}(\mu,{\bm k}_\parallel)\big)^{-1} \nonumber \\
	&~~~~~~~~~~~~\partial_\mu g^S_z(i\omega+\mu,{\bm k}_\parallel)\Big] \nonumber
\end{align}
where $D^S_{a}$ is the surface BC dipole and $d^S_{a}$ its density in momentum space. 

\begin{figure}
  \centering
  \includegraphics[width=\columnwidth]{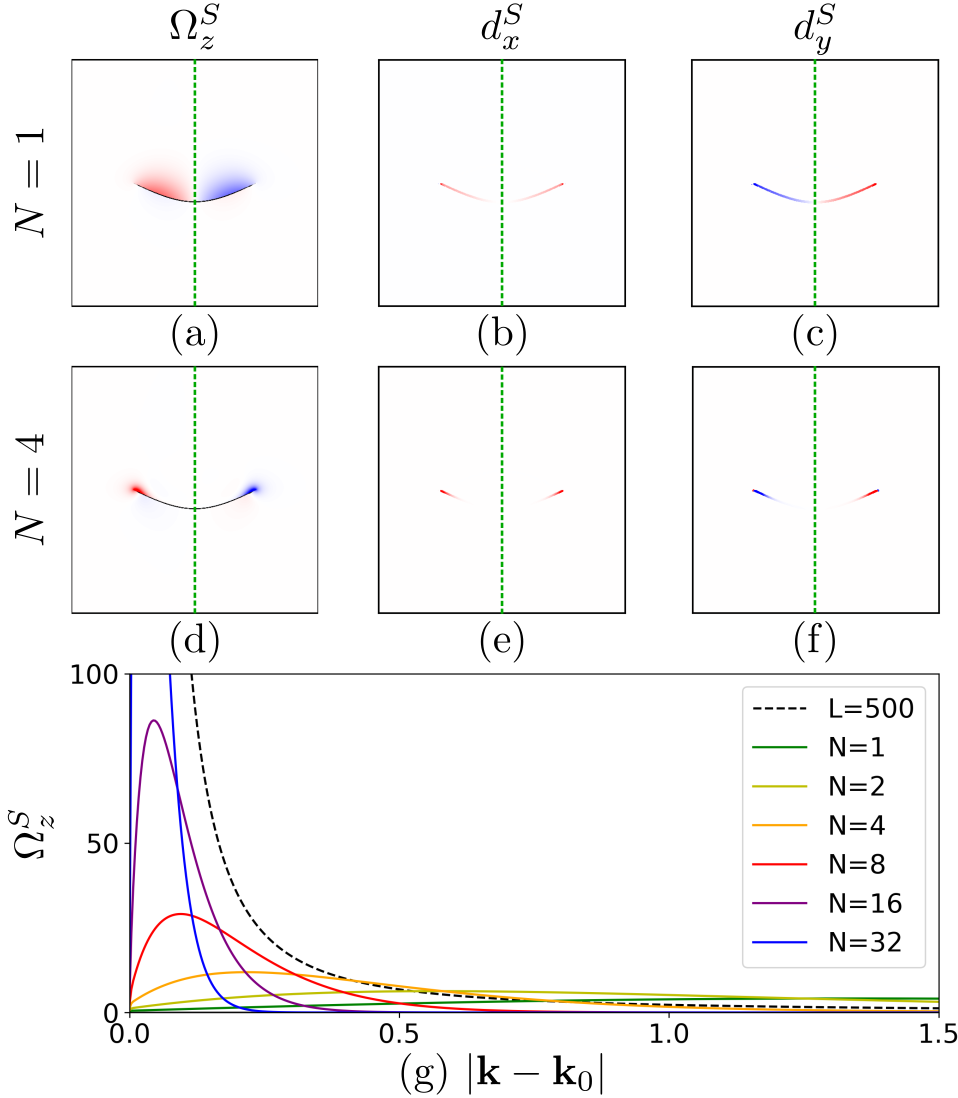}
	\caption{Surface Berry curvature (SBC) $\Omega^s_z$ and SBC dipole density $\vec{d}^S$ of the tilted Weyl cone pair surface (001) associated with a surface layer of thickness (a)-(c) $N=1$ and (d)-(f) $N=4$. The Fermi arcs are marked in black. The SBC dipole $D^S_y$ vanishes due to the dashed green mirror line $M_x$. (g) SBC of the Fermi arc for different $N$ using Green's functions in dependence of the distance $|\mathbf{k}-\mathbf{k}_0|$ from the Weyl node at $\mathbf{k}_0 = (\pi /2,0)$. The dashed line is the result for the surface state of a slab with thickness $L=500$.}
  \label{Fig:BCSGF_Weyl}
\end{figure}

Again, we will take the (001) surface of the Weyl pair model in Eq.\ (\ref{eq:H_pair}). Fig.\ \ref{Fig:BCSGF_Weyl} (a) and (d) show the surface BC associated with the upmost surface layer of thickness $N=1$ and $N=4$, respectively. The second and third columns in Fig.\ \ref{Fig:BCSGF_Weyl} exhibit the components of corresponding BC dipole density which, in line with the present reflection symmetry, yield a finite surface BC dipole only along the mirror line, i.e.\ $D^S_y = 0$.
Fig.\ \ref{Fig:BCSGF_Weyl} (g) presents the BC of the surface state along the Fermi arc parametrized by the distance from the Weyl node at $\vec{k}_0=(\pi/2, 0)$.
To obtain this, we notice that since Eq.\ (\ref{Eq:BC}) gives the surface BC of all occupied bands, the BC of the Fermi arc is the difference between the BC on both sides of it. That its BC is significant can be appreciated in Fig.\ \ref{Fig:BCSGF_Weyl} (a) and even more clearly in Fig.\ \ref{Fig:BCSGF_Weyl} (d). Most importantly both BC and BCD are concentrated near the end of the Fermi arc where it hits the hot-line~\cite{PhysRevLett.127.056601}. As shown in Fig.\ \ref{Fig:BCSGF_Weyl} (g), even though the surface BC is largest close to the projection of the Weyl nodes, it is  convergent. The reason is that the penetration depth of the surface state increases close to a Weyl cone, while we are only calculating the BC of a surface layer of fixed thickness $N$. As the surface layer thickness $N$ is increased, the resulting SBC becomes progressively closer to the analytical $1/k^2$ asymptotic behaviour~\cite{PhysRevLett.127.056601}. Note that, in contrast to the finite slab calculation, the obtained BC includes not only the surface state but also all bulk states in the corresponding $N$ layers. Since the BC of all bands always is zero the bulk bands partially compensate the contribution from the Fermi arc state. Thus, the surface BC obtained from surface GFs is smaller than the pure surface state for a large slab.

\section{Berry curvature mediated transport phenomena}
The Berry curvature of Bloch wave functions has an influence on the velocity of Bloch electrons similar to a magnetic field yielding e.g.\ the anomalous Hall effect. For a Weyl semimetal the BC diverges quadratically at the BC hot-line and might give large responses in experiments. Furthermore, the BC is concentrated at the Weyl nodes and the $k\cdot p$ model will give a good approximation for the following results.

If the WSM has broken TR symmetry it is possible to measure the anomalous Hall effect which is proportional to the integral of the BC over the first Brillouin zone. However, if there is a symmetry like TR or a mirror under which the BC is odd, this integral and the Hall response vanish. Nevertheless, we can have a dipole moment of Berry curvature at the Fermi surface giving rise to a second order anomalous Hall effect yielding a dc and an ac current with twice the applied frequency of the applied electric field\cite{PhysRevLett.115.216806}. In two dimensions this is allowed if the maximal lattice symmetry is a single mirror line.
The Berry curvature dipole of the Fermi arc reads:
\begin{align}
	\vec{D} &= -\int \frac{\dif^2k}{(2\pi)^2}~ \vec{\nabla} f_0 ~\Omega(\kp)\\
	&= -\frac{\chi u_z}{8\pi^2 k_c}\sqrt{1-u_z^2}\frac{\vec{s} 
+ \chi\vec{u}}{1+\chi \vec{s}\cdot\vec{u}}
\end{align}
In the limit $k_c \rightarrow 0$ the BCD diverges due to the diverging BC and for a finite slab the BCD still increases linear in system size $L$. The dipole can not only be measured in the second order non-linear Hall effect $j \propto E^2L$, but also in the non-linear Nernst effect $j \propto (\vec{\nabla}T)^2 L$ and Kerr-rotations and circular dichroism using the effective orbital magnetization $\vec{M} \propto \vec{D}\cdot \vec{E}~\hat{e}_z$.

Another important effect of the BC is the orbital magnetic moment $\vec{m}$ which leads to a deformation of the Fermi surface\cite{PhysRevB.59.14915} and thus, has an influence on transport measurements. If we assume a $\kp$-dependent $\alpha$ this yields for the WSM surface state:
\begin{align}
	\vec{m} &= \text{Im}\br{\bra{\vec{\nabla}\Psi} \times (E-H) \ket{\vec{\nabla}\Psi}}\\
	&= -4 u_z \vec{\nabla}\alpha \times \vec{s}
\end{align}
The orbital magnetic moment does not diverge like the BCD and for a constant $\alpha$ as we have in our models it even vanishes and therefore can be neglected. Regarding the surface state in Eq.\ (\ref{Eq:SurfaceState}), the orbital magnetic moment depends on $\psi(\kp)$ but not on $\lambda(\kp)$.

A further widely known phenomenon is the chiral anomaly that can be seen if a parallel electric and a magnetic field $\vec{E}\parallel\vec{B}$ is applied to a WSM and is caused by the different chiralities of the Weyl nodes. For tilted WSMs there is also a chiral non-linear current\cite{PhysRevB.103.045105}
\begin{align}
	\vec{j} = \int \frac{\dif^2\vec{k}}{(2\pi)^2}~\partial_\epsilon f_0~(\vec{E}\times\vec{\Omega})((\vec{E}\times\vec{\Omega})\cdot (\vec{v}\times\vec{B})).
\end{align}
A natural question is if we can observe such a current caused by the Fermi arc. Since $\vec{\Omega}\perp \vec{v}$ on the surface there is no contribution for $\vec{E}\parallel\vec{B}$. Nevertheless, for $\vec{B} =B \vec{n}$ and $\vec{E}\parallel \vec{v}$ this changes and we get
\begin{align}
	j_x &= -\frac{\sqrt{1-u_z^2}u_z^2}{48\pi^2k_c^3}E^2B \\
	&\propto E^2BL^3
\end{align}
where for simplicity we chose $\vec{n}=\hat{z},~\alpha=\pi/2$, and $\vec{u}_\parallel = 0$. In contrast to the BCD the BC appears squared and the resulting current diverges even faster as $L^3$. Note that it has opposite signs on top and bottom surface and the bulk current vanishes in this case.

If we have a TR symmetric WSM there is still a chance to find a first order Hall-like response. The so-called Magnus Hall conductance\cite{PhysRevLett.123.216802} which is closely related to the BC dipole, yields
\begin{align}
	G_{MH} &= \frac{e^2}{\hbar}\Delta U\int_{v_x>0} \frac{\dif^2 \vec{k}}{(2\pi)^2}~\partial_\epsilon f_0~\Omega \\
	&= \frac{e^2\chi u_z}{8\hbar \pi^2 k_0} \frac{1-u_z^2}{(1+\chi \vec{s}\cdot \vec{u})^2}
\end{align}
for the Fermi arc if the electric field is applied perpendicular to the mirror line and gives rise to a current in $y$-direction along the mirror line that is proportional to the slab thickness $L$.\\

\section{Conclusions and outlook}
We have demonstrated that the surface state of a Weyl semimetal in a lattice model can be derived analytically in a similar manner as the continuum model. The free phase of the self-adjoint extension is fixed in the semi-infinite and finite lattice model resulting in a fully determined shape of the arc and the BC hot-line at which the BC diverges quadratically. We provided a lattice model in which a bulk parameter can change the phase on the surface. A slab calculation confirmed the that Fermi arcs on opposite surfaces hybridize and that the BC dipole grows linearly with the slab thickness $L$. The $1/k^2$ behavior of the surface BC is also revealed using surface Green's functions with different thicknesses of the surface layer. Finally, we showed that the diverging BC results in transport phenomena like the second order Hall effect and the Magnus-Hall conductance diverge linear with the thickness $L$. The second-order non-linear chiral current is even proportional to $L^3$ if electric and magnetic field are perpendicular to each other.

As future direction, the lattice models we have investigated here are well suited for investigating how disorder influences the surface BC and the resulting transport phenomena. Furthermore, finite slabs as well as surface Green's functions may be used to elaborate on the effect of surface reconstructions, which are  relevant from an experimental point of view and may quantitatively alter surface Berry curvatures and their dipoles.

\section*{Acknowledgments}
We thank Jorge I. Facio, Inti Sodemann and Jhih-Shih You for fruitful discussions, Ulrike Nitzsche for technical assistance and the DFG for support through the W\"urzburg-Dresden Cluster of Excellence on Complexity and Topology in Quantum Matter, ct.qmat (EXC 2147, project-id 39085490) and through SFB 1143 (project-id 247310070) project A5. 


%

\end{document}